\documentclass[sigconf,authorversion,nonacm]{acmart}

\usepackage{multirow}
\usepackage{enumitem}
\usepackage{xspace}
\usepackage[commandnameprefix=always]{changes}
\usepackage{verbatim}
\usepackage[normalem]{ulem}
\usepackage{algorithm}
\usepackage{algpseudocode}
\usepackage{float}
\usepackage{soul}

\newcommand{\methodName}{DiffUSEM\xspace}

\renewcommand{\methodName}{DiffEyeSyn\xspace}

\AtBeginDocument{%
  \providecommand\BibTeX{{%
    \normalfont B\kern-0.5em{\scshape i\kern-0.25em b}\kern-0.8em\TeX}}}

\acmSubmissionID{1285}

\copyrightyear{2025}
\acmYear{2025}
\setcopyright{acmlicensed}\acmConference[]{}
\acmBooktitle{}
\acmPrice{15.00}
\acmDOI{}
\acmISBN{}

\acmSubmissionID{1285}

\begin{document}

\title[\methodName: Diffusion-based User-specific Eye Movement Synthesis]{\methodName: Diffusion-based User-specific Eye Movement Synthesis}

\author{Chuhan Jiao}
\affiliation{%
  \institution{University of Stuttgart}
  \city{Stuttgart}
  \country{Germany}
}
\email{chuhan.jiao@vis.uni-stuttgart.de}

\author{Guanhua Zhang}
\affiliation{%
  \institution{University of Stuttgart}
  \city{Stuttgart}
  \country{Germany}
}
\email{guanhua.zhang@vis.uni-stuttgart.de}

\author{Yeonjoo Cho}
\affiliation{%
  \institution{University of Stuttgart}
  \city{Stuttgart}
  \country{Germany}
}
\email{joo9610@gmail.com}

\author{Zhiming Hu}
\affiliation{%
  \institution{University of Stuttgart}
  \city{Stuttgart}
  \country{Germany}
}
\email{zhiming.hu@vis.uni-stuttgart.de}

\author{Andreas Bulling}
\affiliation{%
  \institution{University of Stuttgart}
  \city{Stuttgart}
  \country{Germany}
}
\email{andreas.bulling@vis.uni-stuttgart.de}

\renewcommand{\shortauthors}{Jiao, et al.}

\begin{abstract}

High-frequency gaze data contains more user-specific information than low-frequency data,  promising for various applications. However, existing gaze modelling methods focus on 
low-frequency data, ignoring user-specific subtle eye movements in high-frequency eye movements.
We present \methodName ~-- the first computational method to synthesise eye movements specific to individual users.
The key idea is to consider the user-specific information as a special type of noise in eye movement data.
This perspective reshapes eye movement synthesis into the task of injecting this user-specific noise into any given eye movement sequence. 
We formulate this injection task as a conditional diffusion process in which the synthesis is conditioned on user-specific embeddings extracted from the gaze data using pre-trained models for user authentication. 
We propose user identity guidance ~-- a novel loss function that allows our model to preserve user identity while generating human-like eye movements in the spatial domain.
Experiments on two public datasets show that our synthetic eye movements preserve user-specific characteristics and are more realistic than baseline approaches.
Furthermore, we demonstrate that \methodName can synthesise large-scale gaze data and support various downstream tasks, such as gaze-based user identification.
As such, our work lays the methodological foundations for personalised eye movement synthesis that has significant application potential, such as for character animation, eye movement biometrics, and gaze data imputation.

\end{abstract}

\begin{teaserfigure}
  \includegraphics[width=\columnwidth]{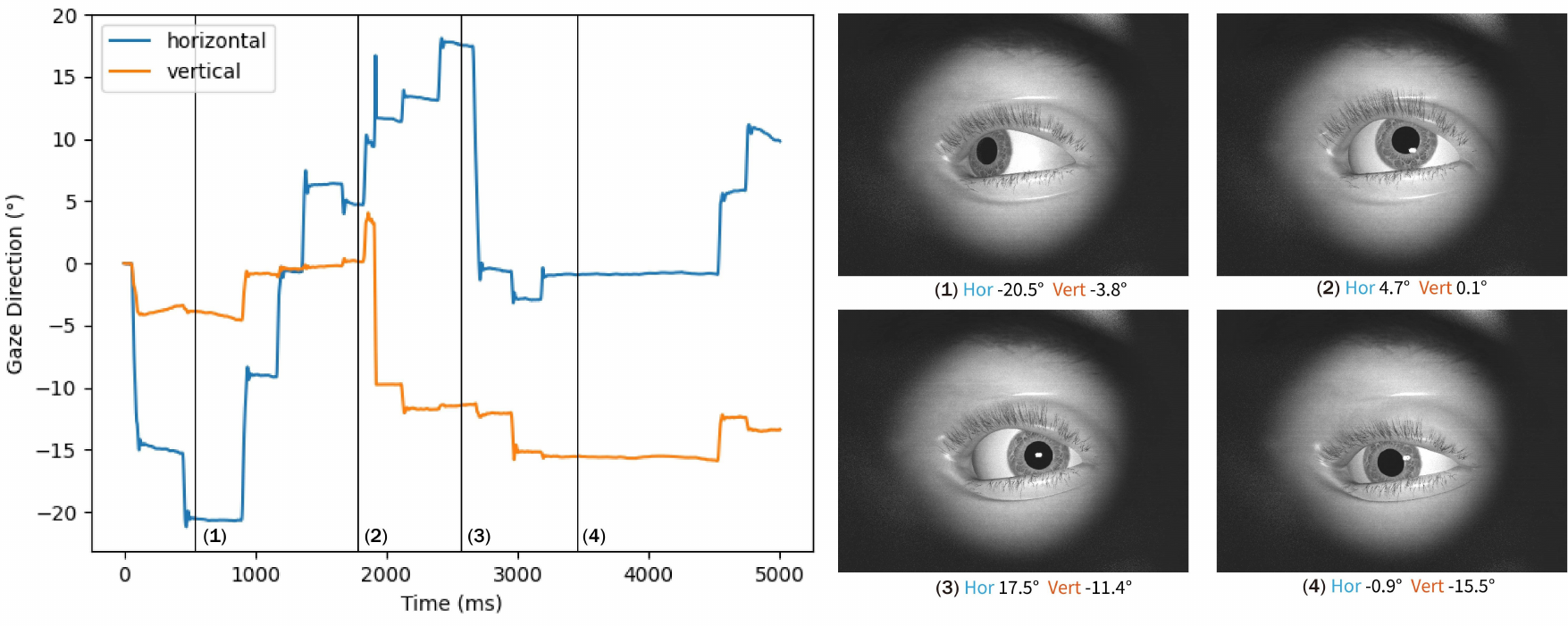}
  \caption{\methodName synthesises user-specific eye movements at 1,000 Hz (Left), demonstrating the potential for applications such as eye movement biometrics and rendering personalised eye animations (Right, rendered using the tool proposed in \cite{swirski2014rendering}, with further examples in the supplementary video). 
  }
  \label{fig:teaser}
\end{teaserfigure}

\maketitle

\section{Introduction}

With significant advances in eye tracking technologies and learning-based gaze estimation~\cite{kassner2014pupil, zhang2017mpiigaze, elfares22_federated}, it has become increasingly convenient to record human eye movements accurately and robustly in everyday settings.
Benefiting from these advances, the human eye movement signal has become a significant modality in the areas of human-computer interaction (HCI) and human-centred computing and has been key to a variety of applications.
Specifically, human eye tracking data has been demonstrated to be highly effective for 1) recognising the activities that a user is performing~\cite{bulling2013eye, hu22_ehtask}; 2) classifying the gender of the user~\cite{sammaknejad2017gender}; 3) estimating the stress level~\cite{huang2016stress} or cognitive load~\cite{fridman2018cognitive} of the user; 4) measuring mind wandering during daily reading scenarios~\cite{faber2018automated}; as well as 5) predicting the actions that a user intends to perform~\cite{lethaus2013comparison}.

While low-resolution eye gaze recordings (typically under 30\,Hz) can fulfil the requirements of some applications like gaze-based activity recognition~\cite{bulling2013eye, hu22_ehtask}, high-resolution eye gaze signals are key to numerous applications such as fast eye movement detection~\cite{leube2017sampling, funke2016eye, dalmaijer2014low} that detects subtle eye movements like microsaccades from high-resolution gaze recordings as well as gaze-based interaction~\cite{duchowski2018gaze,angelopoulos2021event} that reduces the system delay using high-frequency eye gaze data.
More importantly, recent studies in eye tracking research have revealed that the high-frequency eye gaze signals contain rich user-specific information~\cite{holland2012biometric, jiao2023supreyes}.
More importantly, recent studies in eye tracking research have revealed that the high-frequency eye gaze signals contain rich user-specific information~\cite{holland2012biometric, jiao2023supreyes, lohr2022eye, lohr2022eky} than low-frequency ones. 
For example, the performance of using 60-second 30 Hz eye movements falls far short of the performance using 5-second 1,000 Hz monocular eye movements in gaze-based user authentication \cite{lohr2022eye}. 
This indicates that user-specific information within high-frequency eye movement recordings is significant for modelling eye movement behaviours of individual users and has great relevance with various applications, including personalised digital animation (see Figure 1 for example), eye movement biometrics, and gaze data imputation.
Despite the significance of high-frequency eye gaze data, existing works on modelling eye movement behaviours have only focused on low-frequency eye gaze signals.

In this work, we present \textit{\methodName} -- the first computational method to synthesise high-frequency gaze data, including eye movement characteristics specific to individual users.
The key idea of our approach is to inject user-specific information into any given eye movement sequence.
We formulate this injection task as a conditional diffusion process. The synthesis is conditioned on user-specific embeddings extracted from the gaze data using pre-trained user authentication models. 
We propose user identity guidance - a novel loss function that allows our model to preserve user-specific characteristics while generating human-like eye movements in the spatial domain.
Experiments on two public eye movement biometric datasets show that our synthetic eye movements contain user-specific information and are more realistic than existing baseline approaches.
Furthermore, we demonstrate that \methodName can be used to synthesise eye tracking data at scale and for different downstream tasks, such as gaze-based user identification.

\vspace{1em}
\noindent
The specific contributions of our work are four-fold:
\begin{enumerate}

\item We propose \methodName ~-- a novel diffusion-based method that injects user-specific information into any given eye movement sequence to synthesise high-frequency, user-specific eye movement signals.

\item We present user identity guidance ~-- a loss function that can preserve user identity while generating human-like gaze data in the spatial domain. 

\item We report extensive experiments on two public eye movement biometric datasets and demonstrate that our synthetic eye movements contain user-specific information and are more naturalistic than existing eye movement synthesis methods.

\item We demonstrate our method's effectiveness in the sample downstream task of gaze-based user identification.
\end{enumerate}

\section{Related Work}
We discuss related work on 1) eye movement synthesis, 2) eye movement biometrics, and 3) high-resolution eye movement.

\subsection{Eye Movement Synthesis}
Eye movement synthesis research focuses on generating realistic eye movement patterns.
The synthesised data are valuable for various applications, including video rendering of virtual eyes and gaze data augmentation~\cite{duchowski2016eye,lan2022eyesyn}.
Early works primarily originated from general signal processing and computer graphics research and built statistical models.
For example, Yeo et al.~\cite{yeo2012eyecatch} proposed Eyecatch that used Kalman filter to simulate mainly saccades and smooth pursuits.
However, the synthesised trajectories lacked the natural gaze jitter observed in real data.
Duchowski and Jorg~\cite{duchowski2015eye} focused on realistic eye movement rotations based on Donders’ and Listing’s laws.
Based on this model, the authors further considered noise in the eye movements by separating microsaccadic jitter modelled by $1/f$ pink noise and eye tracker noise modelled by white noise to enable more reasonable gaze synthesis~\cite{duchowski2016eye}.
EyeSyn~\cite{lan2022eyesyn} is inspired by psychology, synthesising eye movement data in reading, verbal communication and scene perception, 
modelling different types of eye movements such as fixation, saccade, and smooth pursuit with individual formulas.
Other works attempted to link and generate gaze together with head movements~\cite{ma2009natural, le2012live}.
For instance, Le et al.~\cite{le2012live} nonlinearly mapped features of gaze, eyelid motion and head motion to a high-dimensional feature space and then generated these modalities.

Recent works have started to incorporate deep learning methods to generate eye movement data. 
For example, Fuhl et al.~\cite{fuhl2021fully} leveraged a variational autoencoder (VAE) to simulate eye-tracking data without specific stimuli.
Prasse et al.~\cite{prasse2023sp, Prasse_Improving2024} proposed SP-EyeGAN to synthesise subtle eye movements in-between the known fixation locations.
SUPREYES~\cite{jiao2023supreyes} aimed at gaze super-resolution, i.e., synthesised high-frequency eye movement data upon existing low-frequency data.  
Jiao et al. developed a diffusion-based method, DiffGaze, to synthesise eye movements at 30\,Hz given the $360^{\circ}$ images in virtual reality (VR)~\cite{jiao2024diffgaze}.

However, existing methods have only generated ubiquitous scanpaths or low-frequency eye movements, which neglect the user-specific subtle eye movements manifested in high-frequency data.
In stark contrast, our \methodName is the first approach to synthesise user-specific high-frequency eye movements.
In addition, a key advantage of \methodName is that it can inject these subtle movements into any eye movement sequence, including scanpaths. As such, \methodName can serve as a post-processing method for existing stimulus-driven scanpath prediction approaches. Moreover, our method can be used in various applications, such as eye movement biometrics and personalised eye movement animations.

\subsection{Eye Movement Biometrics}
Eye movement data has emerged as a promising indicator for user biometrics.
Traditional approaches usually extract handcrafted features from eye movements and then feed them into statistical or machine learning algorithms for classification~\cite{bayat2018biometric,cuong2012mel}.
For example, Friedman et al.~\cite{friedman2017method} introduced STAR approach, which used the principal component analysis (PCA) and the intraclass correlation coefficient (ICC).
Holland et al.~\cite{holland2013complex} extracted intricate features, including compound eye hopping and dynamic overshooting, and then built classifiers based on Random Forest (RF) and Support Vector Machines (SVM) for user biometrics.
Li et al.~\cite{li2018biometric} extracted texture features via a multi-channel Gabor wavelet transform and applied SVM for classification.

Recent research mainly uses deep learning methods to extract useful features from eye movement data automatically.
Jager et al.~\cite{jager2020deep} conducted user identification from video-based eye-tracking data by proposing a CNN model.
Makowski et al.~\cite{makowski2020biometric} identified 150 users based on binocular oculomotor signals via proposed convolutional neural networks (CNNs). %
They also proposed DeepEyedentificationLive (DEL) model~\cite{makowski2021deepeyedentificationlive}, composed of two convolutional subnets focusing on ``fast'' (e.g., saccadic) and ``slow'' (e.g., fixational) eye movements, respectively.
Their follow-up work~\cite{makowski2022oculomotoric} finetuned the DEL model to explore drunkenness and fatigue's influences on eye movement biometrics.
Lohr et al.~\cite{lohr2022eky} designed Eye Know You (EKY), a lightweight model that achieved reasonable authentication performance.
This model consisted of exponentially dilated convolutions and only included around 475K parameters.
Later, the authors presented Eye Know You Too (EKYT)~\cite{lohr2022eye}, which applied an end-to-end DenseNet-based CNN architecture and is the state-of-the-art gaze-based user authentication method.

These prior works have observed that high-frequency (1000\,Hz and higher) gaze data contain more user-specific information, promising for various applications, such as user authentication~\cite{lohr2022eye}.
Our work, for the first time, leverages the pretrained user authentication model for user-specific eye movement synthesis and its corresponding evaluation.

\subsection{High-Frequency Eye Movement}
\label{sec:rw-high-resolution}
The peak speed of human eye movement can reach up to 700$^{\circ}/s$ in real life~\cite{duchowski2017eye}.
High-frequency eye movement is hence pivotal in capturing the rapid and subtle movements of human eyes, e.g., microsaccades that typically last for a very short time like 25\,ms~\cite{kothari2020gaze}.
It is also essential for reducing the latency of real-time gaze-based interaction~\cite{angelopoulos2021event}.
Moreover, the subtle eye movements within high-frequency eye movement behaviour are also key to user biometrics as described above.

Traditionally, commercial eye trackers such as EyeLink 1000 Plus\footnote{\url{https://www.sr-research.com/eyelink-1000-plus/}} rely on high-speed cameras to capture high-resolution eye movement data.
However, these cameras are costly in terms of both price and power consumption~\cite{angelopoulos2021event}.
Dynamic vision sensors adaptively sample eye movement and thus offer an alternative solution, but they demand special-purpose hardware and cannot generalise to commonly used eye trackers relying on ordinary cameras~\cite{stoffregen2022event}.
Jiao et al. proposed a deep learning-based method, SUPREYES, that directly converts low-resolution eye movement data into arbitrary high-resolution ones without requiring any additional equipment~\cite{jiao2023supreyes}.
However, this method only focuses on generating high-frequency data that is similar to human gaze data in the spatial domain, ignoring user-specific information. 

Given the advantage of high-resolution eye movement in applications, in this paper, we synthesise high-resolution user-specific eye movement behaviour at a high frequency of 1000\,Hz. 
\section{Background: Denoising Diffusion Probabilistic Models}

Denoising diffusion probabilistic models (DDPMs) \cite{ho2020denoising} is a type of generative model that has demonstrated state-of-the-art performance on various data synthesis tasks, such as image \cite{dhariwal2021diffusion, rombach2021highresolution} and video generation \cite{videoworldsimulators2024}, audio synthesis \cite{kong2020diffwave}, as well as low-frequency gaze sequence generation \cite{jiao2024diffgaze}.
DDPMs consist of two computational processes: A forward process and a reverse process. 
The forward process aims to shift the input data distribution to a normal distribution by adding $T$-step noises.  
The reverse process is where the training and inference happen. The diffusion model is trained in the reverse process to predict how much noise is added to the original data distribution given a noisy distribution and a timestep. Then, starting with a normal distribution, the trained model gradually predicts the noise for $T$-step to denoise the normal distribution back to the original data distribution.

In the forward process, random noise is added according to
\begin{equation}
    q\left(x_t \mid x_0\right)=\mathcal{N}\left(x_t ; \sqrt{\bar{\alpha_t}} x_0,\left(1-\bar{\alpha}_t\right) I\right)
\end{equation} where $x_0$ is the original input data, $x_t$ is the noisy data at timestep $t$, and $\bar{\alpha}_t=\prod_{s=1}^t \alpha_s$ is a hyper-parameter that can be calculated given a predefined noise schedule. The noise schedule indicates the amount of noise should be added at each timestep.

During the reverse process, given noisy data $x_t$, a deep learning model $f_\theta$ is trained to predict the added noise at timestep $t$:
\begin{equation}\label{eq:estimate_noise}
    \hat{\epsilon_t} = {f_\theta}(x_t, t)
\end{equation}
where $\hat{\epsilon_t}$ is the predicted noise.
The training objective of the model is to minimise the difference between the predicted and the actual noise at each timestep:
\begin{equation}\label{eq:loss_simple}
    \mathcal{L}_{simple} = {\left\|{\epsilon_t - \hat{\epsilon_t}}\right\|} 
\end{equation}

Once the model is trained, the generated data is initialised from the normal distribution $x_t$, and realistic data is obtained after $T$ denoising steps. 
The reverse process can be formalised as:
\begin{equation} \label{eq:reverse_ddpm}
    p_\theta\left(x_{t-1} \mid x_t\right)=\mathcal{N}\left(x_{t-1} ; \mu_\theta(x_t, t),\sigma_\theta(x_t, t) I\right)
\end{equation}
where $\mu_\theta(x_t, t)$ can be determined given the predicted noise $\hat{\epsilon}_t$ while $\sigma_\theta(x_t, t)$ is typically a constant value to simplify the optimisation. 
Therefore, each denoising step can be written as:
\begin{equation}
    x_{t-1} = \frac{1}{\sqrt{\alpha_t}}\left(x_t-\frac{1-\alpha_t}{\sqrt{1-\bar{\alpha}_t}} \hat{\epsilon}_t\right) + \sigma_tz, \quad with\ z \sim \mathcal{N}\left(0 ; I\right)
\end{equation}

\section{\methodName}
\begin{figure*}[t]
    \centering
    \includegraphics[width=\textwidth]{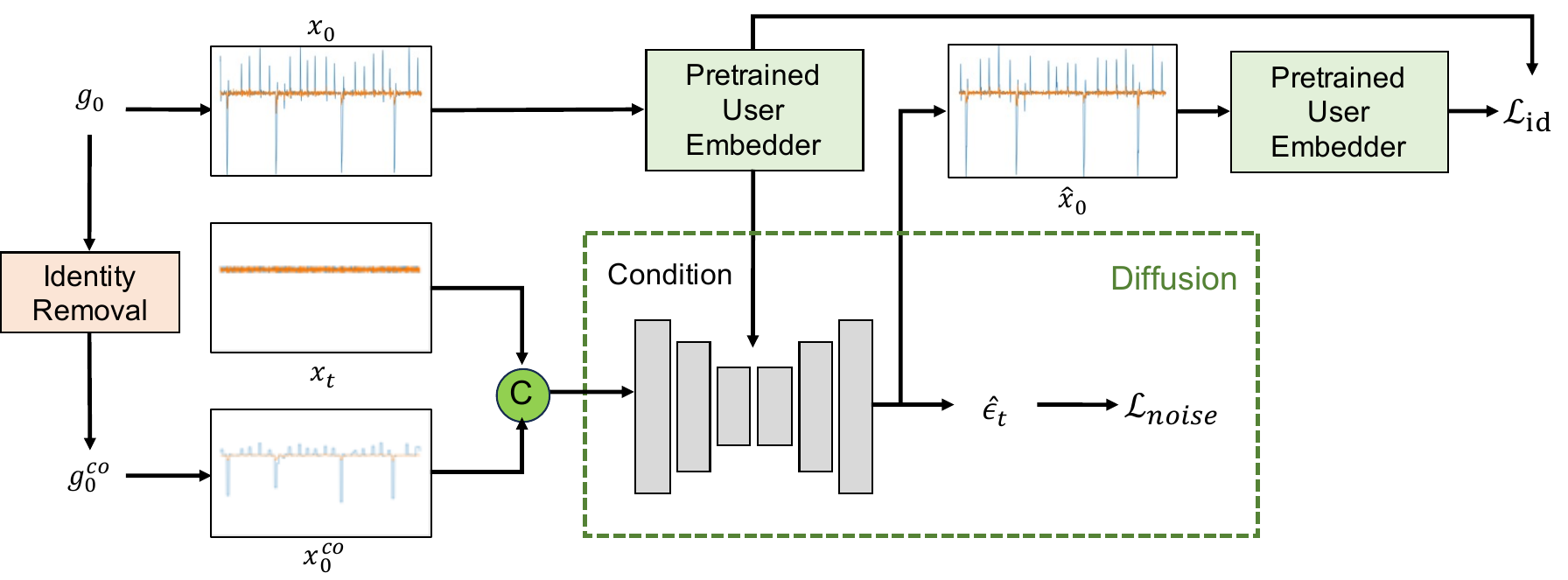}
    \vspace{-0.3cm}
    \caption{Pipeline of training \methodName. \methodName is trained in a self-supervised way. 
    The original eye movement data $g_0$ and its identity removed variant $g_0^{co}$ are converted into velocities $x_0$ and $x_0^{co}$. The goal is to train the diffusion model to inject the removed identity information back into the $x_0^{co}$.
    We use a pretrained user embedder to extract the user-specific embedding from the $x_0$. 
    At each diffusion timestep $t$, given the $x_0^{co}$ and the user embedding as the condition, \methodName predicts the noise $\hat{\epsilon_t}$ that converts the $x_0$ to $x_t$.
    Instead of only optimising \methodName with the normal diffusion loss which minimises the difference between the predicted and ground truth noise, we propose user identity guidance $\mathcal{L}_{id}$ - a novel loss function to constraint the synthesised data contains the given user-specific information. More specifically, we estimate the cleaned eye movement velocity $\hat{x_0}$ by denoising $x_t$ with the predicted noise $\hat{\epsilon_t}$. The proposed user identity guidance maximises the cosine similarity between the embedding of the $\hat{x_0}$ and $x_0$.  
    }
    \label{fig:model-general}
\end{figure*}

\subsection{Problem Definition} \label{sec:problem_definition}

Research on eye movement biometrics (EMB) has indicated that high-frequency eye tracking data contains more user-specific information than low-freqency eye tracking data. \cite{jiao2023supreyes, lohr2022eye, makowski2021deepeyedentificationlive}.
It has been shown that computational models for user identification and authentication, when trained with high-frequency eye movement data, achieve the best performance \cite{lohr2022eye}.
Conversely, models that only had access to low-frequency gaze data (30 Hz or lower) have been deemed unsuitable for practical applications \cite{lohr2022eye, jiao2023supreyes}.
Informed by these findings, and given higher-frequency data captures more subtle eye movements, we consider user-specific information as a special type of noise in eye movement data. Consequently, we define the task of user-specific eye movement synthesis as the process of injecting user-specific noise (subtle eye movements) into any given eye movement sequence.

Since Diffusion models, characterised by their ability to model noise addition and subsequent denoising processes, have shown remarkable success across various domains \cite{ho2020denoising, jiao2024diffgaze, rombach2021highresolution, tashiro2021csdi, kong2020diffwave}, we formulate eye movement synthesis as a diffusion task conditioned on the user-specific information and any given eye movement sequence (see Section \ref{sec:cond_diff} for details), where the user-specific information can be extracted by a pretrained EMB model from a sequence of high-frequency eye movements of the target user. 

Figure \ref{alg:training} outlines the forward process of \methodName, which operates in a self-supervised manner. Initially, a sequence of high-frequency eye movements (e.g., 1,000 Hz) is used to extract the user embedding containing user-specific information via a pretrained EMB model. Subsequently, the user identity information is removed from the high-frequency eye movements (details in Section \ref{sec:implemetation_details}). The diffusion model is then trained to restore the user-specific information on the identity-removed eye movements given the user embedding.

At inference time, given a target-user embedding and a reference sequence of eye movements, the trained \methodName can inject the user-specific information of the target user into the reference eye movements (details in Section \ref{sec:inference}). 

It is worth noting that many EMB models operate on eye movement velocities as input \cite{lohr2022eky, lohr2022eye}. To better adapt EMB models in our method, \methodName synthesises eye movement velocities rather than directly synthesise gaze locations. The gaze locations can subsequently be derived from the synthesised velocities by providing a starting point.

\subsection{Conditional Diffusion Model for Eye Movement Synthesis}\label{sec:cond_diff}
We formulate the task defined in Section \ref{sec:problem_definition} as a conditional diffusion task which is trained in a self-supervised fashion. The training pipeline of the model is shown in Figure \ref{alg:training}. 
The velocity of the ground truth high-frequency gaze data, denoted by $x_0$, serves as the target for our conditional diffusion model, while $x^{co}_0$ represents the velocity of the identity-removed gaze data.
Given a pretrained user authenticator $E$, we extract the user embedding of the ground truth gaze data, denoted as $E(x_0)$.
The $x^{co}_0$ and the $E(x_0)$ are the conditions of \methodName.
The goal of our conditional diffusion model is to estimate the true conditional gaze data distribution $q(x_0 \mid x^{co}_0, E(x_0))$ with a learned distribution $p_\theta(x_0 \mid x^{co}_0, E(x_0))$.
To achieve this, a natural approach is to extend the forward and reverse processes of the DDPM to conditional ones.
Therefore, the predicted noise at timestep $t$ in Equation \ref{eq:estimate_noise} is reformulated as:
\begin{equation}\label{eq:condition_noise_estimation}
    \hat{\epsilon_t} = {f_\theta}(x_t, t \mid (x^{co}_0, E(x_0)))
\end{equation}
The objective is the same as the original DDPM showed in Equation \ref{eq:loss_simple} that minimises the difference between the predicted noise and the ground truth noise:
\begin{equation}
    \mathcal{L}_{noise} = {\left\|{\epsilon_t - \hat{\epsilon_t}}\right\|} 
\end{equation}
Furthermore, the inference process in Equation \ref{eq:reverse_ddpm} is extended to:
\begin{equation} \label{eq:8}
\resizebox{\columnwidth}{!}{$
    p_\theta\left(x_{t-1} \mid x_t\right)=\mathcal{N}\left(x_{t-1} ; \mu_\theta(x_t, t \mid (x^{co}_0, E(x_0))),\sigma_\theta(x_t, t \mid (x^{co}_0, E(x_0))) I\right)
$}
\end{equation}

\subsection{User Identity Guidance}

Training the model with $\mathcal{L}_{noise}$ allows the model to learn the distribution of human eye movements.
However, there is no explicit user identity control in the generation process.
Inspired by the classifier-guided diffusion model in image generation \cite{dhariwal2021diffusion}, we propose a novel \textit{user identity guidance} that leverages the pretrained user authenticator $E$ to allow \methodName to capture the user-specific information of human eye movements in the forward steps. 
At timestep $t$, we obtain $x_t$ by adding a sampled noise $\epsilon_t$ on the target $x_0$.
We then predict the added noise $\hat{\epsilon_t}$ using Equation \ref{eq:condition_noise_estimation}. 
With the predicted noise $\hat{\epsilon_t}$, we can directly estimate the target $\hat{x_0}$ by denoising $x_t$ using:
\begin{equation}
    \hat{x_0} =\frac{x_t-\sqrt{1-\alpha_t} \hat{\epsilon_t}}{\sqrt{\alpha_t}}
\end{equation}
We constrain the embedding of $\hat{x_0}$ and the embedding of $x_0$ to be similar in the embedding space of the user authenticator.
We measure the similarity between embeddings using the cosine similarity, following \cite{lohr2022eye}. The proposed user identity guidance $\mathcal{L}_{id}$ is as follows:
\begin{equation}
    \mathcal{L}_{id} = 1 - \frac{E(x_0) \cdot E(\hat{x_0})}{|E(\hat{x_0})||E(\hat{x_0})|}
\end{equation}
where $E$ is a pretrained user authenticator.

Therefore, the training objective is 
\begin{equation}
    \mathcal{L} = \mathcal{L}_{noise} + \lambda \mathcal{L}_{id}
\end{equation}
In practice, we set the $\mathcal{L}_{noise}$ to be two times important than $\mathcal{L}_{id}$ using:
\begin{equation}
    \mathcal{L} = \frac{\mathcal{L}_{noise}}{\mathcal{L}_{noise}.detach()} + 0.5 \frac{\mathcal{L}_{id}}{\mathcal{L}_{id}.detach()} 
\end{equation}

\subsection{Data Preprocessing} \label{sec:data_preprocessing}
The output of eye trackers is commonly in the form of continuous 2D on-screen gaze locations $(x,y)$ or 3D gaze vectors $(x,y,z)$. However, compared with 3D gaze vectors, the 2D on-screen gaze locations vary across different experimental settings. To increase the generalisability, \methodName uses 3D gaze data in the form of degrees of visual angle, which is represented by the angle of the x-axis and the angle of the y-axis.

\methodName synthesises gaze data at 1,000 Hz. Given a sequence of ground truth eye movements, we first add noise to the eye movements to remove the user identity information (Details about user identity removal are described in Section \ref{sec:implemetation_details}). To obtain the $x_0$ and $x_0^{co}$ for our model, we estimate the velocities (in the unit of degrees per second) of the original eye movements and noisy eye movements in both x-axis and y-axis following \cite{lohr2022eye} using Scipy built-in Savitzky-Golay filter with $windowlength=7$, $order=2$, and $deriv=1$. The invalid values NaNs are inevitable within gaze data due to pupil detection fails or blinks. We assume the eyes are fixated when there is an invalid value. Therefore, all the invalid values in eye movement velocities are replaced by 0. To reduce the noise from eye trackers, we clamp the velocities in the range of $[-1000^{\circ}/s, 1000^{\circ}/s]$. Previous works applied the Z-score standardisation on the velocities before feeding them into the model. We found that the model trained with Z-score standardised data tends to generate the data points that are within the range of $mean \pm std.$ only, ignoring the very fast eye movements. To model more realistic human eye movements, we propose a new data normalisation technique: Velocities are initially rescaled to the range of $[-90, 90]$, then normalised to $[-1, 1]$ through sine transformation.

Algorithm \ref{alg:training} summarises data preprocessing and \methodName training.

\begin{algorithm}[t]
\caption{\methodName Training}\label{alg:training}
\begin{algorithmic}[1]
    \State \textbf{Input}: gaze data $q$, pretrained user authenticator $E$, total diffusion step $T$
    
    \Repeat
        \State \(g_0 \sim q\)
        \State $g^{co}_0 $= IdentityRemoval($g_0$)
        \State $vel_0$ = Savitzky-Golay-filter$(g_0, windowlength=7, order=2, deriv=1)$
        \State $vel_0^{co}$ = Savitzky-Golay-filter$(g_0^{co}, windowlength=7, order=2, deriv=1)$
        \State $vel_0$ = clamp(nan-to-num$(x_0, num=0),$ min=-1000, max= 1000)
        \State $vel_0^{co}$ = clamp(nan-to-num$(x_0, num=0),$ min=-1000, max= 1000)
        \State $x_0$ = $sin(vel_0 / 1000 * 90)$
        \State $x_0^{co}$ = $sin(vel_0^{co} / 1000 * 90)$
        \State \(t \sim \text{Uniform}(\{1,...,T\})\)
        \State \(\epsilon \sim \mathcal{N}(0, I)\)
        \State \( x_t =\mathcal{N}\left( \sqrt{\bar{\alpha_t}} x_0,\left(1-\bar{\alpha}_t\right) I\right)\)
        \State \(\hat{\epsilon_t} = {f_\theta}(x_t, t \mid (x^{co}_0, E(x_0)))\)
        \State \(\hat{x_0} =\frac{\mathbf{x}_t-\sqrt{1-\alpha_t} \hat{\epsilon_t}}{\sqrt{\alpha_t}}\)
        \State Take gradient descent step on
        \State \(\nabla_\theta({\left\|{\epsilon_t - \hat{\epsilon_t}}\right\|}  + \lambda(1 - \frac{E(x_0) \cdot E(\hat{x_0})}{|E(\hat{x_0})||E(\hat{x_0})|})\)
    \Until {converged}
\end{algorithmic}
\end{algorithm}

\subsection{Inference} \label{sec:inference}
A summary of \methodName inference is shown in Algorithm \ref{alg:inference}. At the inference time, \methodName injects the user-specific information into any given eye movements. It is worth noting that \methodName also has the potential for user-specific gaze data imputation. This tasks aim to fill in missing values in existing eye movement sequences. To directly apply \methodName at the inference time, users may utilise any interpolation technique to fill the missing values first, which, alongside a sequence of the target user's eye movements, serves as input to \methodName.

\begin{algorithm}
\caption{\methodName Inference}\label{alg:inference}
\begin{algorithmic}[1]
    \State \textbf{Input}: Any sequence preprocessed reference eye movement velocity $x^{co}$, pretrained user authenticator $E$, a sequence preprocessed eye movement velocity from the target user $x_{target-user}$, total number of denoising steps $T$ 
    \State \text{Sample $x_T \sim \mathcal{N}(0, I)$}
    \State \text{for $t = T$, $T-1$, $\cdot$$\cdot$$\cdot$, 1 do}
         \State \hspace{0.5cm}\(\hat{\epsilon_t} = {f_\theta}(x_t, t \mid (x^{co}, E(x_{target-user})))\)
         \State Compute $\mu_\theta(x_t, t \mid (x^{co}, E(x_{\text{target-user}})))$ 
         \Statex \hspace{1cm} and $\sigma_\theta(x_t, t \mid (x^{co}, E(x_{\text{target-user}})))$ 
         \Statex \hspace{1cm} using $\hat{\epsilon_t}$
         \State \hspace{0.5cm} \text{Sample $x_{t-1} \sim p_\theta\left(x_{t-1} \mid x_t\right)$ using Equation \ref{eq:8}}
    \State \text{end for}
    \State \text{return $x_0$}
        
\end{algorithmic}
\end{algorithm}

\subsection{Architecture}

The architecture of \methodName is built upon DiffWave \cite{kong2020diffwave}. DiffWave is a conditional diffusion model for generating 5-second audio at 22,050 Hz conditioned on mel spectrogram. The bidirectional dilated convolutions (Bi-DilConv) used by DiffWave are memory- and time-efficient for very long input sequences, such as high-frequency eye movements, compared with transformers which have been used in gaze sequence generation \cite{jiao2024diffgaze} and time-series imputation \cite{tashiro2021csdi} diffusion models. 

\begin{figure*}[t]
    \centering
    \includegraphics[width=.9\textwidth]{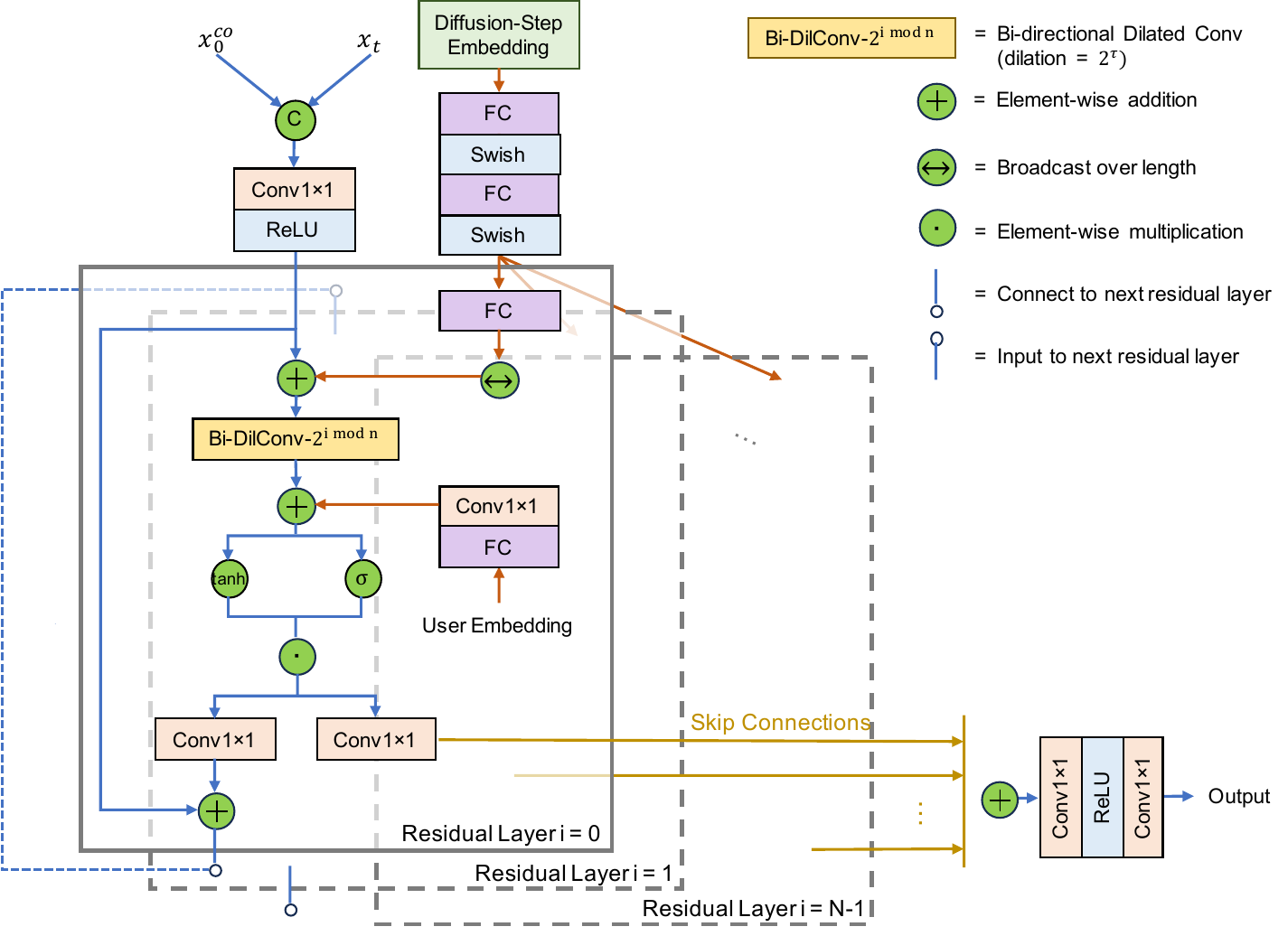}
    \vspace{-0.3cm}
    \caption{Architecture of \methodName. It contains 30 residual layers with  bidirectional dilated convolutions (Bi-DilConv) to ensure the memory- and time-efficient training and inference for generating very high-frequency eye movements.
    }
    \label{fig:model-architecture}
\end{figure*}

Since DiffWave has proved successful in high-frequency realistic audio generation, we only modify a few layers to adapt DiffWave to our settings. Figure \ref{fig:model-architecture} shows \methodName architecture. 
Same as DiffWave, \methodName has a sequence of $30$ residual layers with residual channels $C=64$. Each layer contains a Bi-DilConv layer with kernel size 3. The dilation starts at $1$ at the very first residual layers and doubles at the next layer. The skip connections of all residual layers are summed and contribute to the final prediction.

Since the diffusion step $t$ is one of the inputs and it is a scalar, to help the model distinguish different $t$, it is important to encode $t$ to a high-dimensional space. Following previous studies \cite{vaswani2017attention, kong2020diffwave}, we encode the timestep $t$ as follows:
\begin{equation}
\resizebox{\columnwidth}{!}{$t_{\text {encoding }}=\left[\sin \left(10^{\frac{0 \times 4}{63}} t\right), \cdots, \sin \left(10^{\frac{63 \times 4}{63}} t\right), \cos \left(10^{\frac{0 \times 4}{63}} t\right), \cdots, \cos \left(10^{\frac{63 \times 4}{63}} t\right)\right]$}
\end{equation}
Then the $t_{\text {encoding}}$ is processed by two fully connected layers, each followed by SiLU activations, yielding feature dimensions of $512$.

The main difference to DiffWave architecture is the condition processing. DiffWave employs a mel-spectrogram as its only condition. In contrast, \methodName has two conditions: an observation $x_0^{co}$ and an user embedding $E(x_0)$. Given $x_0$'s length $L$, we merge the noisy input $x_t$ with $x_0^{co}$ along the feature axis. The concatenated tensor with shape $(4, L)$ is then processed by a 1D convolutional layer with the output channel of $C$. In addition, a fully connected layer broadcasts the user embedding $E(x_0)$ to length $L$ before feeding into the residual layer.

Each residual block is equipped with a fully connected layer that broadcasts the processed $t_{\text {encoding}}$ to $C$ channels. The $C$-channel timestep embedding is then added to the aforementioned concatenated tensor. Subsequently, the tensor is upsampled to $2C$ channels by the Bi-DilConv layer. Parallelly, the input user embedding $E(x_0)$ is also upsampled to $2C$ channels by a 1D convolutional layer with kernel size $1$. The upsampled user embedding is added to the upsampled tensor as a bias term for further processing to obtain the predicted noise.

\section{Experiments} \label{sec:exp} 

\subsection{Experimental Setup} \label{sec:implemetation_details}
\textbf{The Selection of Pretrained User Authenticator.} Since there is no explicit approach to evaluate if a sequence of eye movements is from a specific user, inspired by the Fréchet inception distance (FID) score \cite{heusel2017gans} which measures generated image quality using pretrained the Inception V3 model \cite{szegedy2016rethinking}, we propose to use pretained eye movement user authentication model to extract the user embedding from input eye movements and implicity evaluate the generation quality of our method. The performance of \methodName relies on a powerful user authenticator. Therefore, we used the state-of-the-art eye movement user authentication model - EKYT \cite{lohr2022eye} in our experiment. EKYT is a DenseNet-based convolutional neural network, given an input sequence of 5-second 1,000 Hz eye movements the model outputs a 128-dimensional user embedding. In practice, the authors trained four EKYT models through 4-fold cross-validation. The final EKYT user embedding is a 512-dimensional embedding aggregated from the four 128-dimensional vectors. It is worth noting that user authenticators like EKYT generate embedding for input eye movements from any user, including the users shown in training and novel \textit{unseen} users.  

We directly used the pretrained weights provided by the authors. Following the evaluation settings of the user authentication, we calculate the cosine similarity between the EKYT embeddings of the generated eye movements and the ground truth eye movements as the measure of the generated data quality. 

\textbf{Datasets.}
We used two public eye movement biometric datasets, GazeBase dataset \cite{griffith2021gazebase} and JuDo1000 dataset \cite{makowski2020biometric}, in our experiments.

GazeBase is a large-scale dataset of eye movements collected using an Eyelink 1000 eye tracker that records at 1,000 Hz. It has 12,334 eye tracking records from 322 participants, covering different eye movements across nine rounds of recordings over 37 months. Each round has two sessions, and each session has seven visual tasks: fixations (FXS), horizontal (HSS) and random (RAN) saccades, reading (TEX), watching two videos (VD1 and VD2), and playing gaze-driven games (BLG). Further details on this dataset can be found in GazeBase paper \cite{griffith2021gazebase}. To avoid the information leak, we adopted the dataset split described in the EKYT paper, as the EKYT model was trained on this dataset. In particular, the test set comprises recordings from 59 subjects presenting in round 6, while the training set comprises recordings from the remaining 263 participants. The recordings of the BLG task were excluded from the training set. Refer to the EKYT paper \cite{lohr2022eye} for more details. There is \textit{no overlap of users} between the training and test sets. These ensure zero information leakage from the pretrained model and we are able to test \methodName performance on synthesising user-specific eye movements for unseen users during training.

JuDo1000 dataset contains recordings of 150 participants which were collected by an EyeLink Portable Duo eye tracker at 1,000 Hz. Each subject participanted in 4 sessions,  each spaced at least one week apart. The task is similar to the RAN task in GazeBase dataset. We used all the recordings from JuDo1000 for cross-dataset evaluation.

\textbf{Implementation Details.} 
As we chose the pretrained EKYT model to offer user identity guidance, the output of \methodName must match the input format of the EKYT model. Consequently, \methodName produces 5-second 1,000 Hz eye movements, aligning with the EKYT model's requirements. We set the total diffusion steps $T=50$ and we used a linear noise schedule $[1\times10^{-4}, 0.05]$. We trained \methodName on GazeBase training set for $1M$ steps using Adam optimizer with a batch size of 32 and learning rate $2 \times 10^{-4}.$

\textbf{User Identity Removal.} Previous studies in eye movement biometrics \cite{lohr2022eye, lohr2022eky} and gaze data super-resolution \cite{jiao2023supreyes} suggest that low-frequency gaze data at 20-30 Hz performs poorly in gaze-based user identification and authentication. In addition, Jiao et al. \cite{jiao2023supreyes} demonstrate that upsampling the low-frequency gaze data to a high frequency using traditional interpolation methods leads to a performance drop in user identification compared with the original low-frequency data. Based on these findings, we removed user-specific information in the ground truth gaze data by downsampling and interpolation. Specifically, we first downsampled the 1,000 Hz gaze data to 20 Hz and upsampled it back to 1,000 Hz using the previous interpolation. To verify the effectiveness of our identity removal, we calculated the cosine similarity between the EKYT embeddings extracted from the ground truth gaze data and the gaze data after the identity removal on both datasets. The resulting mean cosine similarity values on both datasets, as shown in Table \ref{tab:identity_removal}, are consistently below $0.1$, confirming the effectiveness of our approach in removing user-specific information from the gaze data. Additionally, compared with adding random noise to the eye movements, removing the user-specific information by downsampling and interpolations keeps the low-frequency basis of the eye movements, which simplifies the task for \methodName as it only needs to inject the user-specific subtle movements back to the given eye movement sequence, instead of learning to synthesis both realistic low-frequency and high-frequency eye movements at once. 

\begin{table*}[]
\resizebox{\linewidth}{!}{
\begin{tabular}{lcccccccc}
\toprule
Dataset         & \multicolumn{7}{c}{Task}                                                                                     &               \\ \midrule
                & HSS           & RAN           & TEX           & FXS           & VD1           & VD2           & BLG           & ALL           \\ \cline{2-9} 
GazeBase(Train) & 0.119 ± 0.176 & 0.078 ± 0.197 & 0.047 ± 0.176 & 0.148 ± 0.196 & 0.060 ± 0.198 & 0.057 ± 0.196 & -             & 0.081 ± 0.191 \\
GazeBase(Test)   & 0.135 ± 0.166 & 0.084 ± 0.185 & 0.061 ± 0.174 & 0.153 ± 0.178 & 0.079 ± 0.187 & 0.080 ± 0.189 & 0.030 ± 0.166 & 0.090 ± 0.181 \\
JuDo1000       & -             & -             & -             & -             & -             & -             & -             & 0.091 ± 0.104 \\
\bottomrule
\end{tabular}}
\vspace{2pt}
\footnotesize{$-$ not applicable}
\caption{The cosine similarity between EKYT embeddings of the ground truth gaze data and the identity removed gaze data on GazeBase \cite{griffith2021gazebase} dataset and JuDo1000 \cite{makowski2020biometric} dataset. Cosine similarity being 1 means that two embeddings are exactly the same, while 0 means two embeddings are orthogonal, i.e., not related.}
\label{tab:identity_removal}
\end{table*}

\subsection{User Identity Recovery} \label{sec:user_identity_recovery}
We first conducted an user recovery experiment for \methodName evaluation.

\textbf{Task Definition.} We assume we have the identity-removed eye movements and the user embedding extracted from the ground truth 1,000 Hz eye movements. The objective is to recover the user identity for the identity-removed eye movements. 

We acknowledge that this task's scenario is uncommon in real-world applications because obtaining user embeddings requires access to the ground truth data initially. However, we chose this task because it represents an ideal case, showcasing the upper-bound performance achievable by our method.

\textbf{Experimental Details.} Our experiment was conducted on two datasets: the GazeBase test set, comprising 73,011 non-overlapping 5-second 1,000 Hz gaze sequences, and the JuDo1000 dataset, containing 7,200 non-overlapping 5-second 1,000 Hz gaze sequences. None of the users in these two test sets are shown during training. Given that \methodName is a generative method, we synthesised five samples per gaze sequence to mitigate sampling bias.

\textbf{Evaluation Metrics and Baselines.} We assessed performance using cosine similarity between the EKYT embedding of the synthesised eye movements and that of the ground truth eye movements. We reported the mean cosine-similarity and the standard deviation.

To the best of our knowledge, \methodName is the first method for user-specific eye movement synthesis. We compared it with its ablated version, which was trained without the proposed loss function for user identity guidance. 

We also compared \methodName with a classical signal processing baseline. Since we removed user identity by downsampling eye movements to 20~Hz, this is equivalent to removing the high-frequency components. For the baseline, we applied a Butterworth high-pass filter to extract frequency components above 20~Hz from the ground truth, then added these back to the identity-removed eye movements.

In addition, we benchmarked against SP-EyeGAN \cite{prasse2023sp, Prasse_Improving2024}, an existing eye movement synthesis method. It requires predefined fixation locations, mean, and standard deviation for fixation and saccade durations as input for eye movement synthesis. We trained one SP-EyeGAN model for each task in the GazeBase dataset using their official code. For each eye movement sequence in the GazeBase, we synthesised one eye movement sequence using the corresponding SP-EyeGAN model. For cross-dataset evaluation on the JuDo1000 dataset, we employed the SP-EyeGAN model trained on the RAN task.

Moreover, we established a human baseline to better explain the results. For each dataset, we computed within-user cosine similarity and cross-user cosine similarity. Specifically, within-subject cosine similarity measures how similar a sequence of eye movements is to other sequences from the same user. For each user's sequence of eye movements, we calculated cosine similarities between its EKYT embedding and those of other sequences from the same user, then averaged these values across all users. Achieving an average cosine similarity close to the within-user cosine similarity suggests a method can simulate a user's average gaze behaviour. An average cosine similarity higher than the within-user cosine similarity indicates that the method can simulate gaze behaviour similar to the target eye movement sequence. For the Gazebase dataset, we reported the within-subject cosine similarity of seven tasks and the whole dataset.

Cross-user cosine similarity assesses the similarity between a user's eye movements and those of other users in the dataset. We calculated cosine similarities between a user's sequence of eye movements and those of sequences from other users, then averaged these values across all users. Comparing cross-user cosine similarity with within-user cosine similarity helps verify the efficacy of the pretrained user authenticator (EKYT). A gap between these two values indicates the pretrained user authenticator's ability to differentiate eye movements from different users. Similarly, we reported the cross-user cosine similarity for seven tasks and the whole dataset.

\begin{table*}[!htbp]
\resizebox{\linewidth}{!}{
\begin{tabular}{llcccccccc}
\toprule
Dataset                    & Method                                      & \multicolumn{7}{c}{Task}                                                                                     &                                       \\ \midrule
                           &                                             & HSS           & RAN           & TEX           & FXS           & VD1           & VD2           & BLG           & ALL                                   \\ \hline
                           & Human (within-user)                         & 0.514 ± 0.159 & 0.512 ± 0.154 & 0.506 ± 0.161 & 0.403 ± 0.153 & 0.481 ± 0.158 & 0.491 ± 0.160 & 0.468 ± 0.160 & 0.498 ± 0.160                         \\
                           & Human (cross-user)                          & 0.111 ± 0.168            & 0.110 ± 0.171             & 0.113 ± 0.168             & 0.135 ± 0.159             & 0.117 ± 0.168             & 0.114 ± 0.170             & 0.099 ± 0.168             & 0.113 ± 0.169                         \\ 
                           & SP-EyeGAN \cite{prasse2023sp, Prasse_Improving2024}                                  & 0.114 ± 0.136 & 0.134 ± 0.137 & 0.157 ± 0.142 & 0.106 ± 0.150 & 0.128 ± 0.145 & 0.121 ± 0.144 & 0.135 ± 0.133 & 0.129 ± 0.141                         \\ \cline{2-10}
                           & High-pass Filter (R)                                 & 0.390 ± 0.170 & 0.407 ± 0.178 & 0.320 ± 0.155 & 0.477 ± 0.241 & 0.362 ± 0.178 & 0.362 ± 0.174 & 0.324 ± 0.178 & 0.375 ± 0.180                         \\
                           & \methodName w.o.$\mathcal{L}_{id}$ (R)                                 & 0.404 ± 0.161 & 0.348 ± 0.173 & 0.352 ± 0.169 & 0.176 ± 0.211 & 0.269 ± 0.196 & 0.298 ± 0.193 & 0.407 ± 0.171 & 0.322 ± 0.182                         \\
\multirow{-2}{*}{GazeBase} & \methodName (R)                                 & \textbf{0.820} ± 0.063 & \textbf{0.800} ± 0.064 & \textbf{0.823} ± 0.079 & \textbf{0.705} ± 0.104 & \textbf{0.779} ± 0.081 & \textbf{0.791} ± 0.076 & \textbf{0.791} ± 0.081 & \textbf{0.787} ± 0.078                         \\
\cline{2-10}
                            & High-pass Filter (M)                                 & 0.181 ± 0.148 & 0.178 ± 0.151 & 0.178 ± 0.157 & 0.223 ± 0.143 & 0.190 ± 0.152 & 0.194 ± 0.150 & 0.153 ± 0.145 & 0.182 ± 0.151                         \\
                           
                           & \methodName w.o.$\mathcal{L}_{id}$ (M)                                  & 0.235 ± 0.178 & 0.219 ± 0.179 & 0.209 ± 0.186 & 0.100 ± 0.169 & 0.164 ± 0.181 & 0.171 ± 0.183 & 0.214 ± 0.183 & 0.187 ± 0.179                         \\
 & \methodName (M)                                 & \textbf{0.668} ± 0.101 & \textbf{0.664} ± 0.098 & \textbf{0.696} ± 0.106 & \textbf{0.454} ± 0.144 & \textbf{0.629} ± 0.129 & \textbf{0.644} ± 0.121 & \textbf{0.643} ± 0.110 & \textbf{0.628} ± 0.116                         \\ \midrule
                           & Human (within-user) & -             & -             & -             & -             & -             & -             & -             & 0.685 ± 0.104                         \\
                           & Human (cross-user)                          & -             & -             & -             & -             & -             & -             & -             & 0.367 ± 0.157 \\ 
                           & SP-EyeGAN \cite{prasse2023sp, Prasse_Improving2024}                         & -             & -             & -             & -             & -             & -             & -             & 0.141 ± 0.136 \\ \cline{2-10} 
                           & High-pass Filter (R)                                  & -             & -             & -             & -             & -             & -             & -             & 0.534 ± 0.162                         \\
                           & \methodName w.o.$\mathcal{L}_{id}$ (R)                                  & -             & -             & -             & -             & -             & -             & -             & 0.226 ± 0.126                         \\
\multirow{-4}{*}{JuDo1000} & \methodName (R)                                 & -             & -             & -             & -             & -             & -             & -             & \textbf{0.695} ± 0.072                         \\
\cline{2-10} 
                        & High-pass Filter (M)                                 & -             & -             & -             & -             & -             & -             & -             & 0.424 ± 0.139                         \\
                           & \methodName w.o.$\mathcal{L}_{id}$ (M)                                 & -             & -             & -             & -             & -             & -             & -             & 0.119 ± 0.126                         \\
 & \methodName (M)                                & -             & -             & -             & -             & -             & -             & -             & 0.605 ± 0.091                         \\
\bottomrule
\end{tabular}}
\vspace{2pt}
\footnotesize{$-$ not applicable}
\caption{The cosine similarity between EKYT embeddings of the ground truth eye movement data and synthetic data of different methods in user identity recovery (R) and user identity manipulation (M). The results that are better than the human within-user baseline the models are bolded. Cosine similarity being 1 means that two embeddings are exactly the same, while 0 means two embeddings are orthogonal, i.e., not related.}
\label{tab:user_identity_recovery}
\end{table*}

\textbf{Quantitative Results.} Table \ref{tab:user_identity_recovery} presents the results for both datasets. Across both datasets, the human within-user cosine similarity surpasses the cross-user cosine similarity by a big margin (0.498 vs. 0.113 for GazeBase; 0.685 vs. 0.367 for JuDo1000). This disparity indicates that the chosen pretrained user authenticator effectively distinguishes between different users based on their eye movements. 

In comparison, SP-EyeGAN achieves performance close to the cross-user human baseline on the GazeBase dataset and performs even worse than this baseline on the JuDo1000 dataset, suggesting that it struggles to capture user-specific characteristics in eye movements.

On the other hand, although significantly outperforming SP-EyeGAN, \methodName without user identity guidance performs poorly in user identity recovery on both datasets. Specifically, on the GazeBase dataset, the cosine similarity consistently falls below the human within-user cosine similarity across all tasks by at least 0.06, with an overall cosine similarity 0.176 lower than the within-user cosine similarity. On the JuDo1000 dataset, it even performs worse than the cross-user cosine similarity. These findings highlight the failure of \methodName without the proposed user identity guidance in injecting user-specific information into synthesised eye movements.

The high-pass filter baseline slightly outperforms the ablated \methodName on both datasets but still underperforms relative to the within-user baseline. This finding suggests that user-specific information resides not only in the high-frequency components (above 20~Hz) but also in the low-frequency components of eye movement sequences.

In stark contrast, the full \methodName surpasses the human within-user baseline on both datasets. For the GazeBase datasets, \methodName outperforms the human within-user baseline by a considerable margin (at least 0.3) across all tasks. Despite the task of the JuDo1000 dataset being unseen during training, \methodName achieves slightly better performance than the human within-user baseline (0.695 vs. 0.685), showcasing its strong generalisation ability. These results underscore \methodName's proficiency in reintroducing user-specific subtle eye movements into identity-removed eye movements.

\begin{figure*}[]
    \centering
    \includegraphics[width=0.9\textwidth]{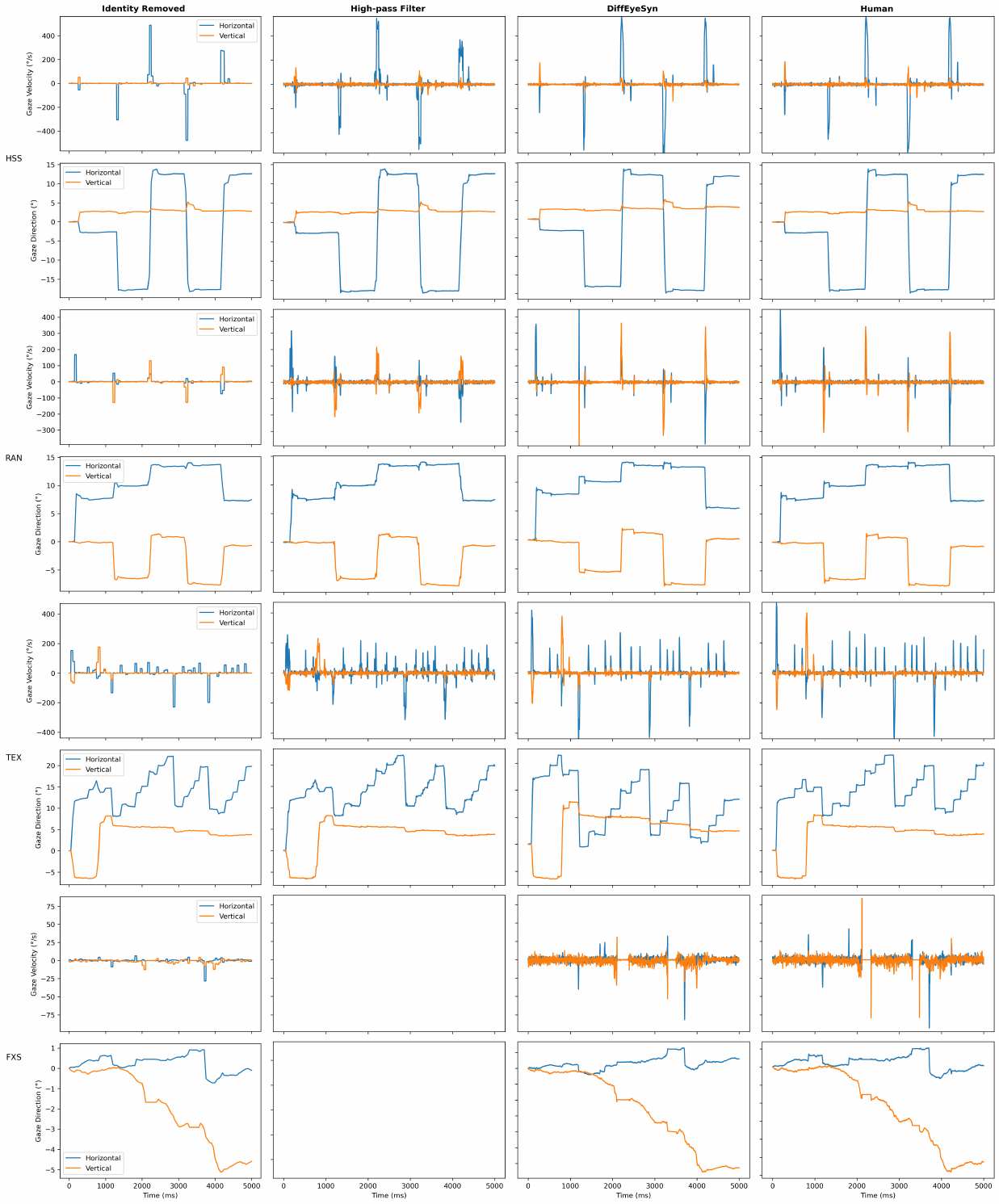}
    \vspace{-0.3cm}
    \caption{Qualitative comparison between the identity removed eye movements, High-pass filter synthetic eye movements, \methodName synthesised eye movements, and ground truth human eye movements in four tasks (HSS, RAN, TEX, FXS) within a 5-second time window. For each sequence of eye movements, we visualise its velocities (above) and gaze direction (below). A figure is empty means the method fails to produce valid results.
    }
    \label{fig:recovery}
\end{figure*}

\textbf{Qualitative Results.} Due to the page limit, we present examples of synthesised eye movements in four different tasks (HSS, RAN, TEX, FXS) in Figure \ref{fig:recovery}. The tasks of TEX, along with others like VD1, VD2, and BLG, represent real-world situations, where choosing one exemplifies the general behavior adequately. Meanwhile, HSS, RAN, and FXS are chosen because they are challenging tasks that with certain types of eye events. More visualisations, including the visualisations of 5-second synthetic eye movements in all seven tasks can be found in the supplementary materials.

Compared with the identity-removed inputs and the synthetic eye movements generated by the high-pass filter, \methodName demonstrates high fidelity in replicating human eye movement velocities across all tasks. In addition, the high-pass filter approach suffers from numerical instability, which can result in invalid output values (see the last two rows in Figure~\ref{fig:recovery}). In stark contrast, \methodName is robust in synthesising user-specific eye movements and is capable of injecting user-specific information into any given sequence of eye movements.

\subsection{User Identity Manipulation}\label{sec:manipulation}

To further evaluate \methodName, we conducted another experiment on user identity manipulation.

\textbf{Task Definition.} In this task, we replicate real-world scenarios where we have a target user and a sequence of their 5-second eye movements. The objective is to generate more user-specific eye movements for this target user. Practically, this involves finding a sequence of eye movements from a different user and incorporating the target user's specific information into these eye movements.

\textbf{Experimental Details.} We recreated the GazeBase test set and JuDo1000 dataset for this task. For each sequence of eye movements in the GazeBase test set, we considered the user of this sequence as the target user. Next, we randomly selected seven other sequences across seven different tasks from seven different users, ensuring that the cosine similarities between the target user's sequence and the selected sequences ranged from 0 to 0.05. This selection criterion ensures that the chosen sequences are dissimilar to the target user's sequence, allowing us to test \methodName in an extreme application scenario. This resulted in our new GazeBase test set for this task containing 511,077 instances of user identity manipulation. We applied the same selection criteria to the JuDo1000 dataset. Since this dataset contains only one task, we randomly selected one sequence from a different user for each sequence of eye movements. Importantly, none of the users in these test sets were included during the training phase. 

Similar to the user identity recovery experiment, we downsampled the selected sequences to 20 Hz to remove the user-specific information.

\textbf{Evaluation Metrics and Baselines.} To evaluate model performance,  we utilised cosine similarity between the EKYT embedding of the synthesised eye movements and that of the target eye movements. Similarly to the user identity recovery experiment detailed in Section \ref{sec:user_identity_recovery}, we established the ablated \methodName, SP-EyeGAN, high-pass filter, human within-user and cross-user cosine similarities as our baselines for comparison. For the high-pass filter baseline in this experiment, we extracted frequency components above 20~Hz from the target user’s eye movements and injected them into the downsampled base sequences.

\textbf{Quantitative Results.} Table \ref{tab:user_identity_recovery} demonstrates the quantitative results. As observed in the user identity recovery results, \methodName without the proposed user identity guidance falls short in user identity manipulation as well. On the GazeBase dataset, its cosine similarity slightly outperforms the cross-user cosine similarity, while on the JuDo1000 dataset, it lags behind the cross-user cosine similarity. In both cases, its cosine similarity is considerably lower than the within-user baselines.

In the Gazebase dataset, the performance of the high-pass filter dropped a lot compared with the performance in the user-identity recovery task; it was only slightly higher than the cross-user baseline. 
Conversely, the full \methodName surpasses the human within-user baselines across all tasks by a minimum of 0.05 and an average of 0.13. This indicates \methodName's capability to inject the user-specific information from the target user's eye movement sequence into random eye movements effectively.

In the cross-dataset evaluation using the JuDo1000 dataset, \methodName achieves a cosine similarity of 0.605, trailing slightly behind the within-user baseline of 0.685. However, it surpasses the cross-user baseline by a margin of 0.238, demonstrating its ability to synthesise eye movement patterns that are meaningfully aligned with the target user's identity. The high-pass filter baseline achieved a cosine similarity of just 0.424, which is 0.181 lower than \methodName, further highlighting the advantage of our proposed method.

\begin{figure*}[]
    \centering
    \includegraphics[width=0.77\textwidth]{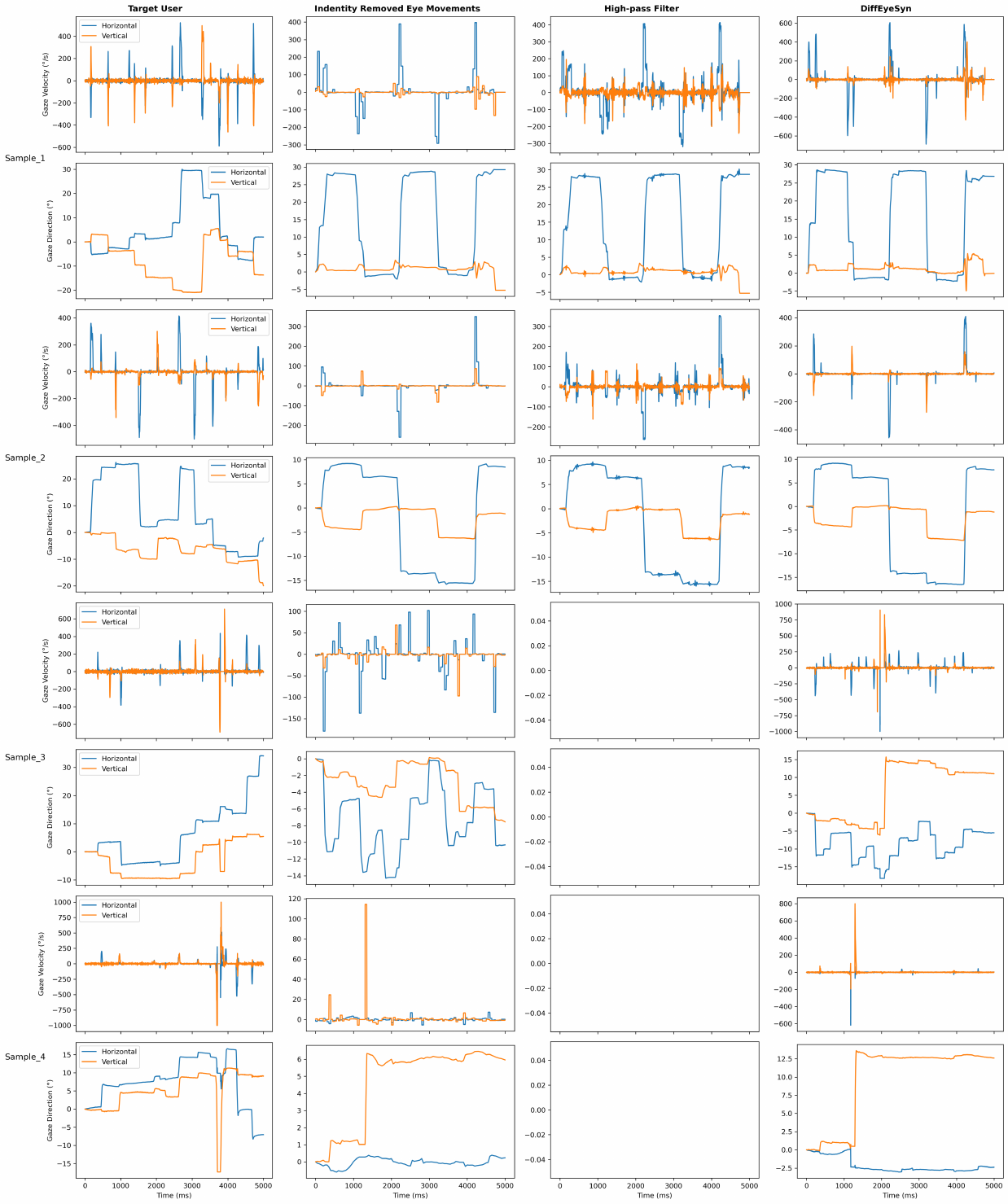}
    \vspace{-0.3cm}
    \caption{Four examples of the user identity manipulation task. Left: the eye movements used to extract the target user embedding. Middle 1: the eye movements from different users that \methodName injects the target user information. Middle 2: High-pass filter synthesised eye movements. Right: \methodName synthesised eye movements. For each sequence of eye movements, we visualise its velocities (above) and gaze direction (below). A figure is empty means the method fails to produce valid results.
    }
    \label{fig:cross_user}
\end{figure*}

\textbf{Qualitative Results.} Figure~\ref{fig:cross_user} presents four examples from the user identity manipulation task. As with the results in Figure~\ref{fig:recovery}, the high-pass filter consistently fails to produce valid outputs in some cases. Furthermore, its generated eye movement velocities appear considerably noisier than those of the target user, leading to unrealistic jitter and instability in the synthesised gaze directions. In contrast, \methodName demonstrates greater robustness and produces synthetic eye movements that are visually more realistic, both in terms of velocity profiles and spatial patterns. Additional examples are provided in the supplementary materials.

\begin{figure*}[]
    \centering
    \includegraphics[width=0.7\textwidth]{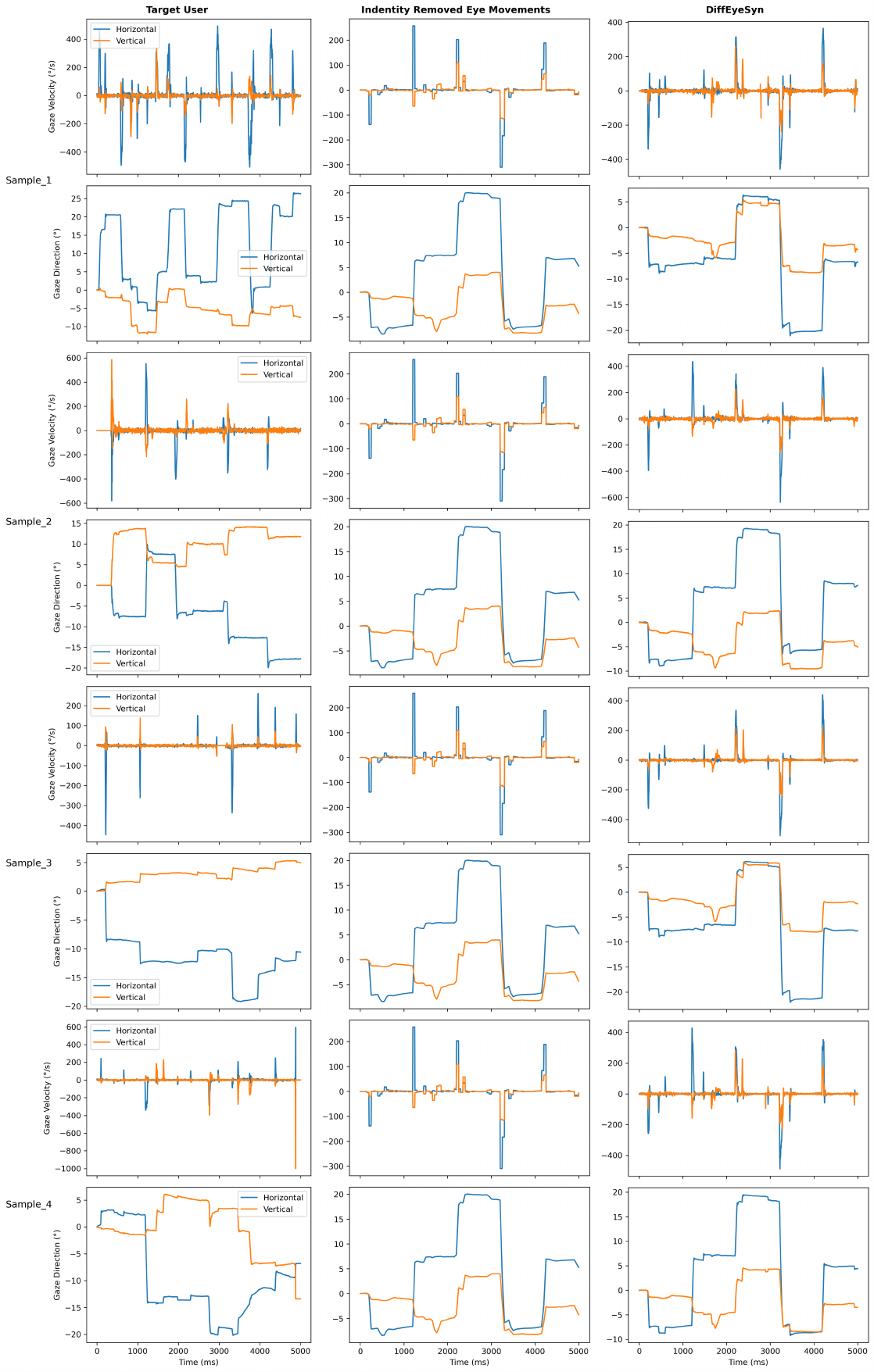}
    \vspace{-0.3cm}
    \caption{Four examples of the user identity manipulation task with the same base eye movement sequence. Left: the eye movements used to extract the target user embedding. Middle: the eye movements that \methodName injects the target user information. Right: \methodName synthesised eye movements. For each sequence of eye movements, we visualise its velocities (above) and gaze direction (below). 
    }
    \label{fig:cross_user_same}
\end{figure*}

To further understand how the base and target eye movements contribute to the synthesis process, we present four additional examples in Figure~\ref{fig:cross_user_same}, where the same base sequence is used with different target user embeddings. All \methodName synthesised results are spatially similar to the base sequences, but show subtle, meaningful differences. This suggests that the base sequence constrains the overall movement range and high-level eye movement structure (e.g., the number, type, and order of events), while the injected user embedding refines finer details, such as producing varying saccade amplitudes at the same saccade locations.

\subsection{Comparison between Synthetic Velocity Distribution and Human Velocity Distribution}
In eye movement synthesis research, prior studies often evaluate synthetic data quality by analysing statistics related to distinct eye events like fixations and saccades \cite{jiao2024diffgaze, prasse2023sp, lan2022eyesyn}. Typically, this involves first detecting various eye events and then examining statistics such as fixation and saccade numbers, along with the velocity distributions of these events. Unlike these works, We opted not to directly compare statistics at the eye event level but instead focus on comparing velocity distributions between synthetic data and real human eye movements. This decision stemmed from our user identity recovery and manipulation experiments, where our emphasis was on reintroducing user-specific information into input eye movements. Given this context, the absolute count of eye events does not vary a lot from actual human eye movements. On the other hand, our method synthesises eye movement velocities and subsequently translates these velocities into gaze directions. Notably, common eye event detection algorithms like I-VT \cite{salvucci2000identifying} and I-DT \cite{salvucci2000identifying} operate based on velocity thresholds or dispersion criteria. Thus, for our synthetic data, identifying eye events is equivalent to detecting them based on eye movement velocities and predefined thresholds. This rationale supports our decision to forego eye event-level statistics in favour of direct velocity distribution comparisons between synthetic and human eye movements.

\textbf{Evaluation Metrics and Baselines.}
Following prior works \cite{prasse2023sp, Prasse_Improving2024, manning1999foundations}, we used the Jensen-Shannon divergence to measure the similarity between two distributions. The Jensen–Shannon divergence ($JS$) is a symmetrised and smoothed version of the Kullback–Leibler divergence ($KL$). Given discrete probability distributions $P$ and $Q$, the $JS$ is defined as:

\begin{equation}
    JS(P \| Q)=\frac{1}{2} KL(P \| M)+\frac{1}{2} KL(Q \| M)
\end{equation}
where $M = \frac{1}{2}(P + Q)$ is the mixed distribution of $P$ and $Q$. The $JS$ is in the range of $[0, 1]$ where greater similarity between distributions is indicated by $JS$ values closer to zero.

We evaluated \methodName against SP-EyeGAN and the high-pass filter baseline using the synthesised data described in Section \ref{sec:exp}.

\textbf{Exprimental Details.} Given the large scale of the two datasets we used, directly comparing distributions from all samples in the dataset becomes infeasible. To address this, we adopted a sampling approach by selecting 1,000,000 velocity samples from each data distribution for distribution comparison. This process was repeated ten times to mitigate sampling bias, and we recorded the average $JS$ value (marked as ALL in Table \ref{tab:velocity}). 

Additionally, we employed the I-VT algorithm~\cite{salvucci2000identifying} to identify fixations and saccades. Specifically, eye movements with velocities below $100^\circ/\text{s}$ were classified as fixations, while those exceeding $300^\circ/\text{s}$ were categorised as saccades. In line with our approach for computing the average JS divergence between the velocity distributions of all synthetic samples and the human ground truth, we also calculated and reported the JS divergence between the velocity distributions of synthetic fixations and real fixations, as well as between synthetic and human saccades.

This additional evaluation is motivated by the fact that the majority of human eye movements fall within the fixation category, which generally involves slower velocities. Consequently, a method might achieve favourable JS scores across all samples by simply modelling low-velocity fixations well, while failing to capture the more dynamic saccadic behaviour. By examining fixations and saccades separately, we gain a clearer picture of the method’s ability to model the full spectrum of eye movement velocities.

To compute these distributions, we used the \texttt{histogram} function from \texttt{NumPy}. As velocities were clipped at a maximum of $1000^\circ/\text{s}$, we selected \texttt{bins = 500} for all samples, \texttt{bins = 350} for saccades, and \texttt{bins = 50} for fixations. This ensured a consistent bin width of 2 across all histograms, allowing for direct comparison between JS divergence scores.

\textbf{Quantitative Results.} The results are summarised in Table~\ref{tab:velocity}. Across both tasks and datasets, the velocity distributions of synthetic eye movements generated by \methodName consistently yield the lowest $JS$ divergence scores when compared with those produced by SP-EyeGAN and the high-pass filter baseline. This indicates that \methodName generates synthetic data that most closely approximates the velocity distribution of real human eye movements.

Notably, \methodName retains this performance advantage not only when considering the full set of samples but also when fixations and saccades are evaluated separately. This suggests that \methodName is capable of faithfully modelling both slow and rapid eye movement behaviours—an essential quality for synthesising realistic and nuanced gaze data. In contrast, the high-pass filter exhibits markedly higher divergence across all categories, reflecting its limited capacity to replicate human-like eye movement dynamics, which is consistent with the qualitative observations shown in Figure~\ref{fig:cross_user}. While SP-EyeGAN achieves comparable performance to \methodName on fixation data, it performs noticeably worse in modelling saccadic movements, further highlighting the strengths of our approach in capturing fast eye movement behaviour.

\begin{table*}[]
\begin{tabular}{llccc}
\toprule
Dataset     & Synthetic Eye Movements                                      & \multicolumn{3}{l}{JS divergence ($\times 10^{-2}$) $\downarrow$}                \\
           &                                                                    & \multicolumn{1}{l}{Fixation} & Saccade       & ALL           \\ \hline
\multirow{5}{*}{GazeBase} & SP-EyeGAN \cite{prasse2023sp, Prasse_Improving2024}    & 1.28 & 7.98 & 1.59  \\ \cline{2-5}
& Hiph-pass Filter (R)    & 15.68 & 4.68 & 15.43  \\
& \methodName (R)    & \ul{1.10} & \ul{0.35} & \ul{1.05}  \\
\cline{2-5}
& Hiph-pass Filter (M)    & 18.44 & 3.94 & 18.17 \\
& \methodName (M)    & \textbf{1.01} & \textbf{0.27} & \textbf{1.00}  \\\midrule
\multirow{5}{*}{JuDo1000} & SP-EyeGAN \cite{prasse2023sp, Prasse_Improving2024}    & 5.79 & 2.87 & 4.63  \\ \cline{2-5}
& Hiph-pass Filter (R)    & 6.66 & 4.35 & 6.69  \\
& \methodName (R)    & \textbf{2.26} & \textbf{0.10}& \textbf{2.17}  \\
\cline{2-5}
& Hiph-pass Filter (M)    & 7.25 & 8.81 & 7.43  \\
& \methodName (M)    & \ul{2.69} & \ul{0.36} & \ul{2.59} \\\bottomrule

\end{tabular}
\caption{Jensen-Shannon divergence between velocity distribution of ground truth human eye movements and velocity distribution of synthetic eye movements of different methods from experiments mentioned in Section \ref{sec:user_identity_recovery} and \ref{sec:manipulation}. The best results are bolded and the second best are underlined. R: user identity recovery, M: user identity manipulation.}
\label{tab:velocity}
\end{table*}

\subsection{Augmenting Gaze-based User Identification}
To further evaluate the effectiveness of \methodName in capturing user-specific information, we conducted an additional gaze-based user identification experiment involving 59 users from the GazeBase test set. In this experiment, we augmented the training data by incorporating synthetic eye movements generated by \methodName. If the synthetic data fails to preserve user-specific characteristics, the identification model trained on the augmented dataset would be expected to underperform compared to a model trained solely on the original data, as previously reported in \cite{jiao2023supreyes}.

\textbf{Experimental Details.} For each user in the GazeBase test set, we randomly selected 50\% eye movement trajectories for training, and the rest for testing. Additionally, we created a \methodName-augmented training set by synthesising seven new samples for each original training instance, as described in Section \ref{sec:manipulation}.

We employed the EKYT model for user identification, extending it with a final linear classification layer. The models were trained separately on the original human data and the \methodName-augmented dataset. Both models were trained for 100 epochs using the Adam optimiser, with a learning rate of 0.001 and a batch size of 128. To minimise training variability, we ran each model five times with different random seeds. In line with prior work \cite{jiao2023supreyes}, we reported the average classification accuracy and equal error rate (EER) as evaluation metrics.

\begin{table}[!htbp]
\centering
\begin{tabular}{lll}
Training data & Acc. (\%) $\uparrow$ & EER (\%) $\downarrow$ \\ \hline
Human & 57.77 ± 2.98 & 1.72 ± 0.19 \\
Augmented & \textbf{59.63 ± 1.72} & \textbf{1.70 ± 0.20} \\ \hline
\end{tabular}
\caption{User identification results of the same model trained on the Gazebase (Human) dataset and the \methodName augmented dataset. The best results are bolded. Acc: classification accuracy.}
\label{tab:downstream}
\end{table}

Table~\ref{tab:downstream} shows that the model trained on the dataset augmented with \methodName achieves superior performance across both evaluation metrics. This finding suggests that \methodName effectively captures user-specific features during synthesis, thereby enhancing the utility of existing gaze datasets for downstream tasks.

\section{Discussion}

\subsection{Importance of User-Specific Eye Movement Synthesis and Applications}
Previous research has primarily focused on modelling human eye movement behaviour at low frequencies, typically below 30 Hz \cite{jiao2024diffgaze, wang23_tvcg}, often overlooking the user-specific subtle movements embedded in high-frequency eye tracking data. In contrast, we propose the first computational approach to modelling user-specific eye movement behaviour. User-specific eye movement synthesis is a promising research field due to its wide-ranging application potential. 

One notable application of user-specific eye movement synthesis lies in training user authentication or identification models. Collecting a large amount of eye movement data from diverse users is time-consuming. For instance, the creation of the GazeBase dataset \cite{griffith2021gazebase} required years of data collection efforts. With a robust user-specific eye movement synthesis model like \methodName, obtaining training data at scale becomes more feasible. Researchers and practitioners can acquire a limited number of eye tracking recordings from subjects to extract their unique embeddings, subsequently leveraging \methodName to synthesise user-specific data efficiently. Notably, \methodName is designed to synthesise 5-second eye movements, but it is not confined to this duration. Since \methodName produces eye movement velocities, cohesive eye movement signals with any duration can be generated by patching and clipping several 5-second eye movement velocities. This flexibility enhances the practicality of \methodName for various applications, allowing for scalable and customised data generation.

Furthermore, user-specific information plays an important role in tasks such as gaze data imputation. While interpolation methods have achieved good time-series metrics, such as mean absolute error, in filling missing data, they lack the ability to provide user-specific solutions \cite{jiao2023supreyes}. \methodName directly addresses these gaps. By incorporating observed eye movement sequences and target user embeddings, \methodName complements interpolation methods by offering personalised, high-frequency eye movement data. For example, in Section \ref{sec:user_identity_recovery} and \ref{sec:manipulation}, we demonstrated \methodName's ability in an extreme user-specific gaze data super-resolution and imputation task (upsampling 20 Hz eye movements to 1,000 Hz). \methodName synthesised eye movements achieve high cosine similarities in both tasks. However, we acknowledge that current outcomes (Figure \ref{fig:cross_user_same}) show that while \methodName effectively captures user-specific information of the target user, the synthesised eye movements may exhibit minor spatial variations from the input data, depending on the target eye movement sequence. While this is acceptable for tasks like user-specific eye movement synthesis, where subtle user-specific eye movements are key, it may limit performance in spatially-sensitive applications. Since many authentication models rely on velocity features rather than spatial ones, we did not directly compare \methodName with interpolation methods or super-resolution techniques that only optimise for spatial accuracy.

Beyond these core applications, \methodName also supports tasks like personalised eye movement animation. For example, it can be used to animate virtual characters with natural, user-specific gaze behaviour at high frame rates (see Figure~\ref{fig:teaser} and supplementary video).

\subsection{ Performance Upper Bound}\label{sec:dis_pretrained}

\methodName is built upon the assumption that there is a pretrained strong user authentication model that can distinguish if two sequences of eye movements belong to the same user robustly. We picked the state-of-the-art user authentication model EKYT \cite{lohr2022eye} over other methods for training and evaluating \methodName. We acknowledge that the performance of \methodName heavily relies on the pretrained user authenticator. It is worth noting that, our proposed method is not bound to the EKYT model, the pretrained user authenticator is just a plug-in to \methodName. We hope there will be a large, powerful pretrained user authentication model for eye movements, like CLIP \cite{radford2021learning} and Stable Diffusion \cite{rombach2021highresolution} in computer vision, to boost \methodName performance in synthesising user-specific eye movements in the near future.

\subsection{Evaluating Synthetic Eye Movements}

Traditional scanpath metrics are not suitable for evaluating our task. Prior work has shown that such metrics perform poorly in assessing synthetic eye movements and fail to align with expert visual judgement~\cite{jiao2024diffgaze}. Moreover, scanpath prediction differs fundamentally from our objective. While scanpath models aim to predict fixation locations and durations for a given visual stimulus, \methodName focuses on generating subtle, individual-specific gaze patterns independent from visual stimulus. A key advantage of \methodName is its ability to inject these subtle movements into any existing eye movement sequence, including scanpaths. As such, these tasks are complementary: \methodName can serve as a post-processing module to personalise the outputs of scanpath prediction models.

Table~\ref{tab:user_identity_recovery} shows that adding high-frequency components from the ground truth to an identity-removed signal improves performance but still falls short of \methodName in recovering user-specific information. This highlights that user identity is not tied solely to event types or frequency components, but rather embedded in the fine-grained dynamics of gaze behaviour—patterns likely shaped by each individual's physiological eye system.
As user-specific information is encoded in these subtle variations, existing quantitative metrics are insufficient for evaluation. At present, the most reliable approach is to assess downstream tasks where user-specificity is critical. Gaze-based user authentication and identification are prime examples of such tasks. Accordingly, we propose using a pretrained user authenticator to evaluate the ability of \methodName to preserve identity-related information and apply \methodName in a downstream task of augmenting gaze-based user identification.

We adopt the EKYT model~\cite{lohr2022eye}, a state-of-the-art method for gaze-based user authentication, both for training and evaluation. While using the same model across both stages risks overfitting to its particular biases, this practice is common in the vision community. For example, DiffusionCLIP~\cite{kim2022diffusionclip} and other recent works~\cite{kim2023dense, yang2023paint, li2023diffusion} use the CLIP model in both training objectives and evaluation metrics. However, we acknowledge that the model trained in this way might be overfitting to the limitations of the selected pretrained model. We anticipate that stronger and more general gaze encoders will emerge in future, allowing for more robust evaluation of user-specific synthesis methods.

In addition to downstream performance, we follow prior work~\cite{prasse2023sp, Prasse_Improving2024} and report Jensen–Shannon ($JS$) divergence between the velocity distributions of synthetic and real eye movements. We also provide qualitative results in Section~\ref{sec:exp} and in the supplementary materials, enabling visual inspection of realism. Our overall goal is to simulate realistic human eye movements. \methodName represents an important first step towards this goal and demonstrates state-of-the-art performance in capturing velocity distributions—an aspect that is largely overlooked by existing scanpath methods.

\subsection{Limitation and Future Work}
We evaluated our method on the GazeBase \cite{griffith2021gazebase} and JuDo1000 \cite{makowski2020biometric} datasets. These datasets have similar quality characteristics, such as being collected using an Eyelink stationary eye tracker at a very high frequency (1,000 Hz) with participants' heads fixed. This similarity allows for a robust evaluation of synthetic user-specific eye movements.
In recent years, eye tracking-enabled head-mounted devices, such as Apple Vision Pro and Meta Quest, have become popular commercial products. Eye movement-based user authentication has significant potential to become a regular unlock method for such headsets. However, head-mounted devices typically employ low-frequency eye trackers, and the recorded eye movements include both head and eye movements.
Using \methodName to synthesise eye movements for training user authentication systems for head-mounted devices requires additional efforts. This includes retraining the user authenticator on a virtual reality (VR) or augmented reality (AR) dataset and subsequently using this retrained user authenticator to fine-tune \methodName. We will investigate this in future work.

\subsection{Ethic Statement}
All the data used in this work are from public datasets, eliminating the concerns about participant privacy during our analysis.

With respect to the proposed model, since we focus on adapting and optimising diffusion models, we do not think this research has any ethical disadvantages beyond those of diffusion models.
Main concerns of such generative models, in the hands of bad actors, could be used to synthesise fake eye movement data or attack user authentication systems.
That said, these tools may also benefit the growth and accessibility for the creative or biometric industry.

\section{Conclusion}

This paper presented \methodName, the first computational approach for user-specific eye movement synthesis. We conducted extensive experiments to evaluate the performance of \methodName in user identity recovery and user identity manipulation, demonstrating \methodName can synthesise realistic user-specific eye movement data that contains user-specific characteristics. In particular, we also demonstrated that \methodName can be directly used in gaze-based user identification. We believe that \methodName sets the basic groundwork for user-specific eye movement synthesis that has significant application potential, such as for personalised character animation, eye movement biometrics, and user-specific gaze imputation.

\bibliographystyle{ACM-Reference-Format}
\bibliography{sample-base}

\end{document}